\documentclass[letterpaper, 10 pt, journal, twoside]{IEEEtran}  

\IEEEoverridecommandlockouts                              
\usepackage{graphicx} 
\usepackage{amsmath} 
\usepackage{amssymb}  
\usepackage[usenames, dvipsnames]{color}
\usepackage{siunitx}
\usepackage{cases}
\usepackage{lipsum}
\usepackage{color}
\usepackage{cite}
\usepackage[ruled,vlined]{algorithm2e}
\usepackage{diagbox} 
\usepackage{tabu}
\usepackage{array}
\usepackage{multirow}
\usepackage{booktabs}
\usepackage{makecell}
\usepackage{empheq}
\usepackage{threeparttable}

\newcolumntype{P}[1]{>{\centering\arraybackslash}p{#1}}
\newcolumntype{M}[1]{>{\centering\arraybackslash}m{#1}}

\newcommand{\revision}[1]{\textcolor{black}{#1}} 

\title{
VesNet-RL: Simulation-based Reinforcement Learning for Real-World US Probe Navigation
}

\author{Yuan Bi*$^{1}$, Zhongliang Jiang*$^{1}$, Yuan Gao$^{1,2}$, Thomas Wendler$^{1}$,\\Angelos Karlas$^{3,4}$ and Nassir Navab$^{1,5}$, \textit{Fellow, IEEE} 

\thanks{Manuscript received: February, 24, 2022; Accepted April, 18, 2022.}
\thanks{This paper was recommended for publication by Editor Jessica Burgner-Kahrs upon evaluation of the Associate Editor and Reviewers' comments. (Yuan Bi and Zhongliang Jiang contributed equally to this work). (\textit{Corresponding author:} Zhongliang jiang and Yuan Bi: zl.jiang@tum.de and yuan.bi@tum.de)}
\thanks{$^{1}$Y. Bi, Z. Jiang, Y. Gao, T. Wendler, and N. Navab are with the Chair for Computer-Aided Medical Procedures and Augmented Reality, Technical University of Munich, Boltzmannstr. 3, 85748 Garching bei M\"unchen, Germany      }%
\thanks{$^{2}$Y. Gao is with the Institute of Medical Robotics, Shanghai Jiao Tong University, Shanghai, China.} 
\thanks{$^{3,4}$A. Karlas is with the Institute of Biological and Medical Imaging, Helmholtz Zentrum München, Neuherberg, Germany, and also the Department for Vascular and Endovascular Surgery, rechts der Isar University Hospital, Technical University of Munich, Germany.} 
\thanks{$^{5}$N. Navab is also with the Laboratory for Computer-Aided Medical Procedures, Johns Hopkins University, Baltimore, USA.}

\thanks{Digital Object Identifier (DOI): see top of this page.}
}

\begin{document}

\maketitle

\begin{abstract}
Ultrasound (US) is one of the most common medical imaging modalities since it is radiation-free, low-cost, and real-time. In freehand US examinations, sonographers often navigate a US probe to visualize standard examination planes with rich diagnostic information. However, reproducibility and stability of the resulting images often suffer from intra- and inter-operator variation. Reinforcement learning (RL), as an interaction-based learning method, has demonstrated its effectiveness in visual navigating tasks; however, RL is limited in terms of generalization. To address this challenge, we propose a simulation-based RL framework for real-world navigation of US probes towards the standard longitudinal views of vessels. A UNet is used to provide binary masks from US images; thereby, the RL agent trained on simulated binary vessel images can be applied in real scenarios without further training. To accurately characterize actual states, a multi-modality state representation structure is introduced to facilitate the understanding of environments. Moreover, considering the characteristics of vessels, a novel standard view recognition approach based on the minimum bounding rectangle is proposed to terminate the searching process. To evaluate the effectiveness of the proposed method, the trained policy is validated virtually on 3D volumes of a volunteer's in-vivo carotid artery, and physically on custom-designed gel phantoms using robotic US. The results demonstrate that proposed approach can effectively and accurately navigate the probe towards the longitudinal view of vessels.

\end{abstract}

\markboth{IEEE Robotics and Automation Letters. Preprint Version. Accepted April, 2022}
{Bi \MakeLowercase{\textit{et al.}}: VesNet-RL: Simulation-based Reinforcement Learning for Real-World US Probe Navigation}




\bstctlcite{IEEEexample:BSTcontrol}
 \section{Introduction}


\begin{figure}[ht!]
\centering
\includegraphics[width=0.48\textwidth]{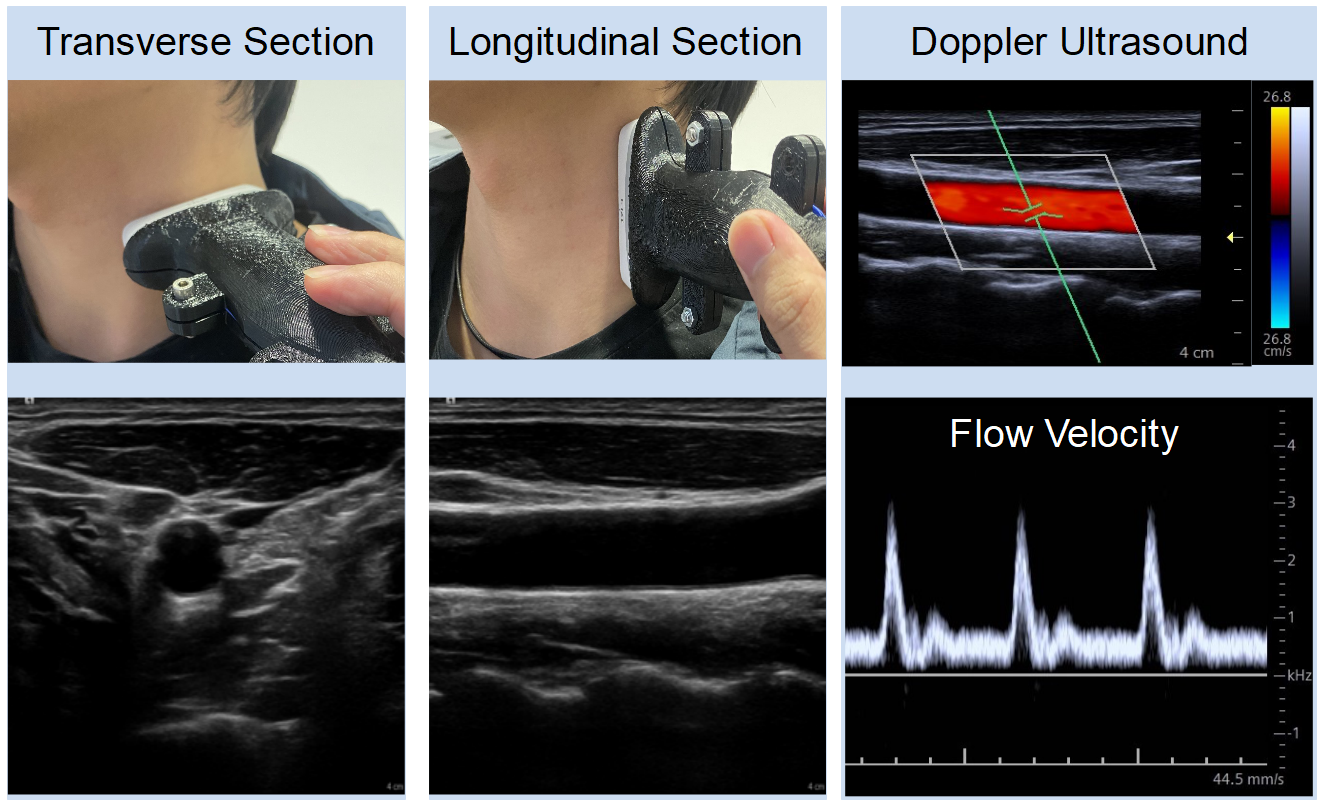}
\caption{Illustration of standard planes navigation task, from transverse section to longitudinal section on a representative carotid artery where the flow velocity of the blood can be measured by doppler imaging.}
\label{Fig_standard_plane}
\end{figure}


\IEEEPARstart{I}{N} the field of medical imaging, ultrasound (US) is one of the most popular diagnostic tools for medical examinations of internal organs. Compared to computed tomography (CT) and magnetic resonance imaging (MRI) examinations, US is real-time, low cost and radiation free~\cite{hoskins2019diagnostic}. 
For vascular medicine, in particular, US plays \revision{a critical} role in everyday practice, namely, for the diagnostics, image-guided interventions and therapy assessment of diseases. In carotid ultrasonography, the optimal acquisition of the longitudinal view of the carotid artery (see Fig.~\ref{Fig_standard_plane}) is required for evaluation of the intima-medial thickness (IMT)~\cite{nezu2015carotid}, the plaque morphology~\cite{spence2002carotid}, or the peak systolic velocity of the blood over plaques~\cite{inzitari2000causes}. As for real-time US-guided femoral arterial access, longitudinal views of the target vessel provide a clear visualization of the needle path and the real-time guidance of the guidewire in the vessel of interest~\cite{seto2010real}.

\par
Such planes are often defined as standard planes in US examinations. To properly display the standard planes, sonographers often need to be trained for \revision{a} few years to gain the necessary anatomical and clinical knowledge. 
However, since the quality of US imaging highly depends on the level of the operator's experience, the conventional freehand US often suffers from low reproducibility (both intra- and inter-operator)~\cite{kanters_reproducibility_1997}. \revision{Furthermore}, force-induced deformation also degrades the imaging quality by introducing non-homogeneous deformation images produced~\cite{jiang2021deformation}.


\par

\subsection{Robotic US}
\revision{Due to the superior performance in accuracy, and repeatability, robotic technologies have been employed to develop a robotic US system (RUSS) to overcome the limitation of operator-variation and further improve the clinical acceptance of US modality.} Since US imaging quality is highly related to the contact force, Pierrot~\emph{et al.} employed a 6-DoF robotic manipulator with \revision{a} compliant controller to maintain a constant force between the patient skin and the US probe~\cite{pierrot1999hippocrate}.
Besides, Hennersperger~\emph{et al.} proposed a workflow to realize autonomous US scans based on imaging registrations~\cite{hennersperger2016towards}.
To optimize probe orientation, Jiang~\emph{et al.} proposed a method to estimate the normal direction of the contact surface based on the force measured at the tip of US probe~\cite{jiang2020automatic,jiang2019automatic}. Yet, the aforementioned work is not \revision{aimed at determining} an optimal view based on the live feed of the US.


\par
Benefited from the development of machine learning, some learning-based approaches have been introduced to address complex recognition and exploring tasks for surgical robotics~\cite{lu2020learning,lu2021toward} or autonomous driving~\cite{zou2021kam}. 
Specific to the task of US standard views recognition, Baumgartner~\emph{et al.} proposed SonoNet to \revision{assist} clinicians \revision{in identifying} the fetal standard planes in real-time during mid-pregnancy US examinations~\cite{baumgartner2017sononet}. In order to provide guidance to sonographers for standard planes \revision{navigation}, Droste~\emph{et al.} used an imitation learning-based system to predict the next action and the final position of the standard view~\cite{droste2020automatic}. However, due to the nature of imitation learning, the demonstrations cannot include all the state space. \revision{Hence}, it is necessary to allow the agent interact with the environment and update its learned policy based on the feedback~\cite{hussein2017imitation}. 
\revision{Reinforcement learning (RL), on the other hand, provides a unique and alternative solution, since the foundation of RL is based on interaction with the environment.}

\subsection{Reinforcement Learning for RUSS}
\par
RL has been proved to be reliable to solve complex decision making and exploration problems and has achieved human-level performance in various scenarios, including virtual environments like Atari games~\cite{mnih2015human}, real-life applications such as robotic grasping, and indoor navigation tasks~\cite{zeng2018learning,zhu2017target}. To exploit the potential of RL in the medical field, Alansary~\emph{et al.} implemented a deep Q-learning based RL framework to locate the standard planes in brain and cardiac MRI volumes~\cite{alansary2019evaluating}. 
Regarding RL applications on RUSS, Hase~\emph{et al.} trained a DQL agent based on the US images recorded from volunteers' spine to guide a US probe to visualize sacrum. The effectiveness of their proposed method was \revision{demonstrated in a virtual environment using unseen data}~\cite{hase2020ultrasound}. Nonetheless, only 2-DoFs translational movements were considered, \revision{implying that} a good orientation \revision{initialization} is \revision{required} for its success. Li~\emph{et al.} then \revision{took a} step forward \revision{by proposing} a DQL framework \revision{that accounts for} all 6-DoFs while constraining the movements of the agent to the patient surface~\cite{li2021autonomous}. Similar to~\cite{hase2020ultrasound}, they trained and tested their agent in a virtual environment built by real spine US images. The relatively \revision{unsatisfactory} performance on the unseen dataset limits its \revision{applicability} in \revision{real-world} scenarios. Unlike~\cite{hase2020ultrasound,li2021autonomous}, who used US images as state representations to navigate the US probe, Guo~\emph{et al.} \revision{attempted to infer} the information of US images and force from scene images, and used the scene as state representation to train \revision{an} RL agent. They used proximal policy optimization algorithms~\cite{ning2021autonomic} and demonstrate the performance on both phantom and humans. 

\par
\revision{Due to the nature of RL, it requires a large number of training episodes, which hinders the possibility to train an RL agent directly with a robot in a real-world scenario}~\cite{zhao2020sim}. Data collection is \revision{a difficult} task and \revision{the generalizability of the trained agent is hampered by the limited and biased training data set.} \revision{Thus, bridging the simulation-reality gap remains a challenge for the community.}



\subsection{Proposed \revision{Solution}}

\begin{figure*}[ht!]
\centering
\includegraphics[width=0.85\textwidth]{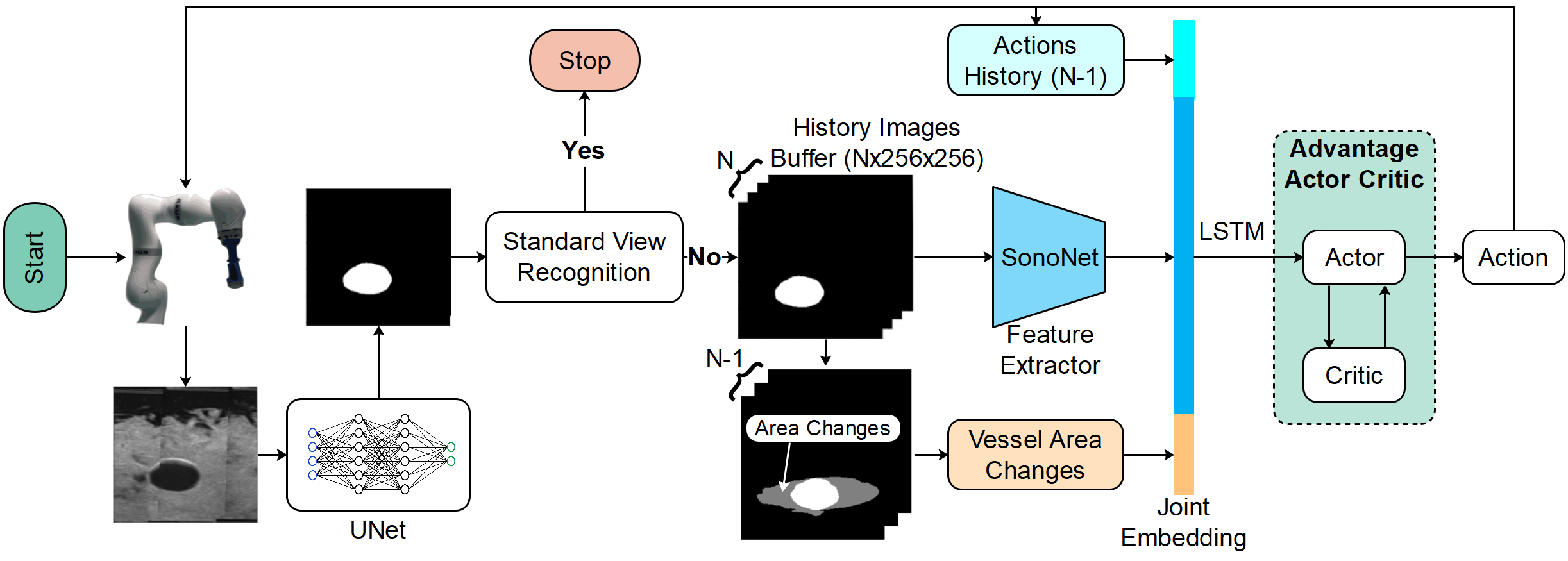}
\caption{Schematic overview of VesNet-RL.}
\label{Fig_overview}
\end{figure*}

\par
\revision{To address these issues, we proposed an RL-based framework, namely VesNet-RL, to perform US standard plane searching for vascular anatomies. We present a method with high generalization ability by first applying a UNet to segment US images and then running RL on the segmentation results. To eliminate the ambiguity caused by the symmetry of vessel, multi-modality information is involved to create a comprehensive state representation. Since the longitudinal view of the vessel appears as a rectangle across the whole US frame, this view can be easily identified by using the minimum area rectangle of the vessel area in US image. The following are the paper’s main contributions:}
\begin{itemize}
  \item An advantage actor critic (A2C) deep RL agent is trained based on the real-time observations to navigate a US probe to the longitudinal view of a vessel. Considering the whole procedure can be interpreted as a partially observable Markov decision process (POMDP), a long short-term memory (LSTM) cell is implemented to exploit the useful information from sequential data.
  \item In order to make the trained model transferable from a simulation environment to real scenarios and even to other similar vascular applications, we used a UNet to segment the vessel area from the US images before using it as a state representation. To create a comprehensive representation of the probe state, a multi-modality state representation is proposed, including a sequence of consecutive segmented US images, the action history, and sequential changes of the segmented area.
  \item A novel standard plane recognition method is introduced based on the minimum area bounding rectangle of the segmented area to estimate the real-time vascular diameter and identify the longitudinal view of vessels.
\end{itemize}
Finally, the proposed \revision{VesNet-RL} is validated both virtually on a volunteer's carotid and physically on a phantom
\footnote{The code: https://github.com/yuan-12138/VesNet-RL}  \footnote{The video: https://www.youtube.com/watch?v=bzCO07Hquj8}.



\section{Method}
The proposed \revision{VesNet-RL} (see Fig.~\ref{Fig_overview}) is based on the standard structure of actor-critic RL agents. The actor-network generates the action based on the observation of the current timestamp, \revision{whereas} the critic-network estimates the preference of the current state. Based on the feedback reward, the network is updated using policy gradient methods. To improve the generalization of the model, the US images are first segmented using a UNet so that the irrelevant background information is erased and the network can focus on the meaningful elements, i.e., vessels. Thereby, the training can be done on the simulated binary images (see Fig.~\ref{Fig_virtual_env}a), where real B-mode images are not \revision{required}. \revision{This speeds up the training process and expands the diversity of the training set.} \revision{An RL agent’s performance is determined by its ability to accurately estimate} its relative position to the goal based on current observations. For such purpose, the features extracted from the history images are concatenated with the previous actions and sequential changes of the segmented vessel area to create a comprehensive representation of the state. \revision{Area changes of the vessel denote the size difference of the segmented vessels in US frames between each step.} In addition, \revision{an} LSTM cell is applied to extract potential crucial information from historical data. The searching process will be stopped when the minimum area bounding rectangle condition is triggered in the standard view recognition module. 


\subsection{RL-Based US Standard Plane Acquisition}
\par
\revision{A Markov decision process (MDP) is a standard RL architecture that includes} a set of states $\mathcal{S}$, a set of actions $\mathcal{A}$, a transition dynamics $\mathcal{T}(s_{t+1}|s_t,a_t)$, a reward function $\mathcal{R}(s_t)$, and a discount factor $\gamma \in [0,1]$. The actual state of the US probe is not directly observable in the US standard plane acquisition task, resulting in a partially observable MDP (POMDP).

\par
\subsubsection{Action Space}\label{sec:Action_Space} In our system, all the translational and rotational movements are performed in the end-effector coordinate frame (CF) \revision{allowing the trained agent to be used in a variety of standard plane acquisition setups.} In comparison, the performance of an RL agent using \revision{an} action space in base CF is dependent on the initial layout of the target object. 
Translational movements along the x- and y-axis of the end-effector CF with $5 mm$ step size and the rotational movements around the z-axis of the end-effector CF with $10^\circ$ step size make up our action space, which is associated with three DoFs of the probe. The probe's movements are eventually restricted to a plane (object surface), defined as an operation surface (OS), where the searching task takes place.

\par
\subsubsection{State and Observations} The actual state of our agent is defined as the relative position between the probe and the target. Because the real state cannot be directly measured, observations such as US images together with actions history and segmented area changes are used as states ($o_t$) to estimate the actual state.

\par
\subsubsection{Reward} Since the goal of RL is to train an agent that can execute optimal policy to maximize the expected accumulated rewards, the reward function actually provides guidance to the agent and determines the objective of the learned policy. In our case, the reward should motivate the agent to locate the vessels's largest longitudinal section. \revision{The size of the segmented vessel area is also considered in the reward design, rather than just the distance to the goal.} The translational distance to the standard view position can be defined as:
\begin{equation}\label{distance}
d_t=\frac{\|(p_t-p_{l_1})\times(p_t-p_{l_2})\|_2}{\|p_{l_2}-p_{l_1}\|_2}
\end{equation}
where $p_t$ is the current position of the probe in the base CF, $p_{l_1}$ and $p_{l_2}$ are two points on the projected vessel centerline in the OS. Because the vessel's centerline in our setup can be approximated as a straight line, $d_t$ actually measures the distance between the probe and the projected vessel centerline in the OS.

\par
Afterward, the score of the current state \revision{in relation to the distance to the} goal is given by:
\begin{equation}\label{score_dis}
\nu_{dis,t}=1-\frac{d_t}{d_{max}}
\end{equation}
where $\nu_{dis,t}\in[0,1]$ and $d_{max}$ is the maximum distance between the probe and the projected vessel centerline in the current virtual environment. This is determined by the location of the vessel and the size of the virtual environment.

\par
The score of the current state is also related to the size of the segmented vessel area in the current US frame:
\begin{equation}\label{score_vessel}
\nu_{ves,t}=\frac{D_t}{D_{max}}
\end{equation}
where $D_t$ is the size of the segmented vessel of the US images (size: $256\times256$) in pixel, and $D_{max}$ denotes the largest segmented area in this simulation environment, $\nu_{ves,t}\in[0,1]$.

\par
The overall score $\nu_t$ of the current state is defined as:
\begin{equation}\label{score}
\nu_t=\mu_{dis}\nu_{dis,t}+\mu_{ves}\nu_{ves,t}
\end{equation}
where $\mu_{dis}$ and $\mu_{ves}$ are the weights of $\nu_{dis,t}$ and $\nu_{ves,t}$, $\mu_{dis}+\mu_{ves}=1$ ,and $\nu_t\in[0,1]$. Here the two weights are set to $\mu_{dis}=0.2$ and $\mu_{ves}=0.8$. Then, the reward function is given by:
\begin{equation}\label{reward}
r_t = \left\{ \begin{array}{ll}
-0.2, & if\: D_t<D_{th}; \\ 
1, & if\:near\:the\:goal\\
5, & if\:reaching\:the\:goal\\
\nu_t-\nu_{t-1}, & otherwise.
\end{array}
\right.
\end{equation}
where $D_{th}$ is \revision{the confirmation} threshold for the vessel's existence. If the size of the segmented area is smaller than $D_{th}$, which, we assume, means that there is no vessel in the US image, the agent will be punished by $-0.2$. A positive ($+1$) reward is assigned to the agent if it is near the goal ($\nu_t>0.9$), while a higher reward ($+5$) is gained by the agent when it reaches the goal ($\nu_t>0.95$). Otherwise, the reward is given by the change of the score ($\nu_t-\nu_{t-1}\in[-1,1]$).
\par
\subsubsection{Advantage Actor Critic} The RL is basically intended to find an optimal policy $\pi^*(a_t|s_t)$, that maximizes future reward at each step~\cite{arulkumaran2017deep}. By estimating the policy $\pi$ with an actor-network, parameterized by $\theta$, the goal can be refactored as maximizing the objective function $J(\theta)$ representing the sum of the reward of the trajectories $\tau$ selected by $\pi_\theta$.
The optimisation of the actor network is then done by gradient ascent $\theta \leftarrow \theta +\eta \nabla_\theta J(\theta)$. The gradient of the objective function takes the form:
\begin{equation}\label{J_theta_gradient}
\nabla_\theta J(\theta)=\mathbb{E}_{\tau\sim \pi_\theta}[\nabla_\theta \log \pi_\theta (s_t,a_t)\psi_t]
\end{equation}
where $s_t$ and $a_t$ are state and action in the trajectory $\tau$. $\psi_t$ can have different designs, which distinguishes different policy gradient approaches. For A2C, it is defined by:
\begin{equation}\label{a2c}
\psi_t=R_t-V_\omega=\sum_{k=0}^{n-1} \gamma^k r_{t+k+1} + \gamma^n V_{\omega}(s_{t+n+1})-V_{\omega}(s_t)
\end{equation}
where $V_{\omega}$ is the critic network parameterized by $\omega$, which is a function estimator of the value function.
The critic network is updated to minimize the mean square error between the estimated and real values.
\par
\begin{figure}[ht!]
\centering
\includegraphics[width=0.44\textwidth]{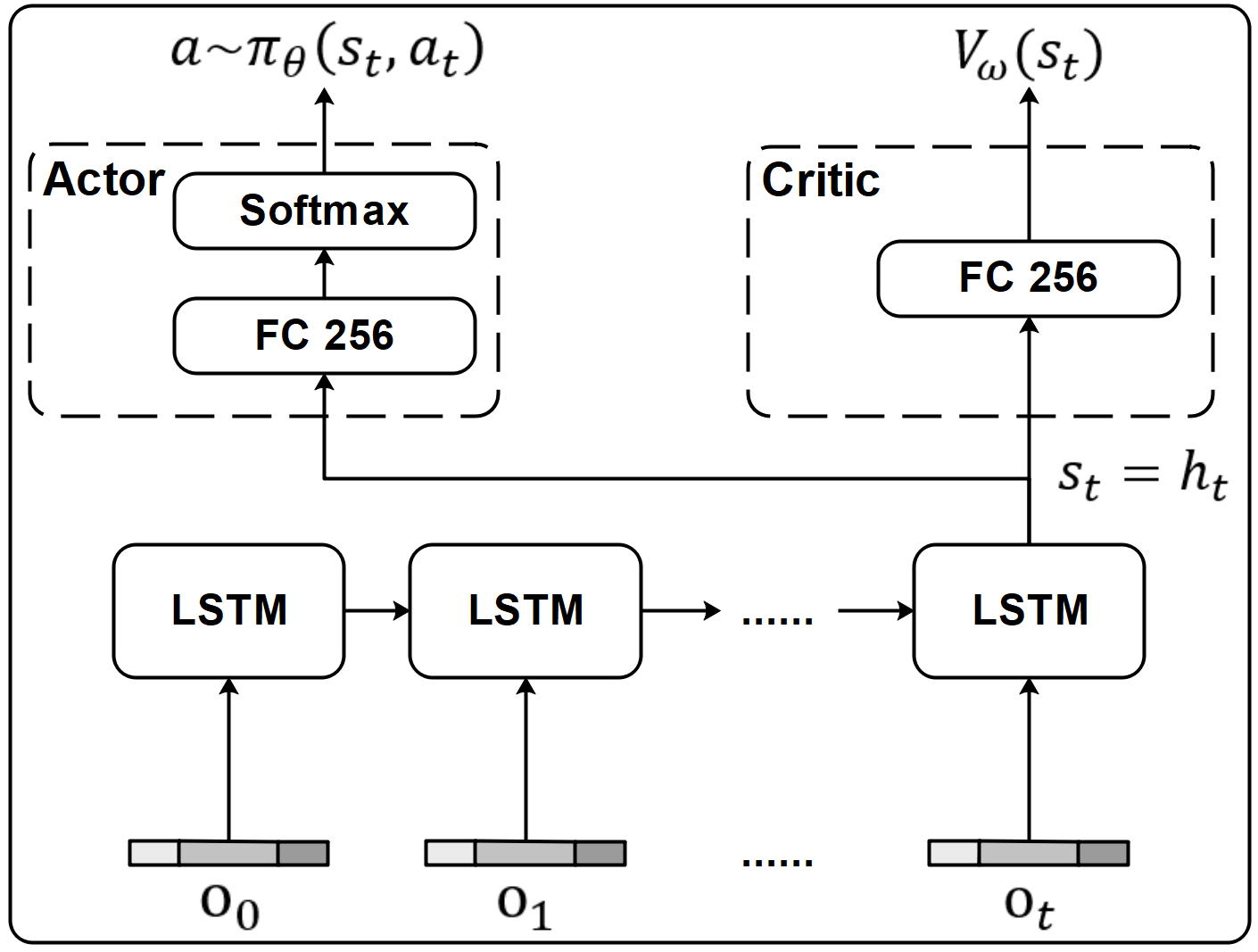}
\caption{Illustrations of actor critic architecture with LSTM cell.}
\label{Fig_A2C}
\end{figure}

\revision{As previously mentioned, the entire process in our problem setting is a POMDP. To deal with the uncertainty introduced by POMDP,} an LSTM cell is implemented to make full use of the sequential information~\cite{mnih2016asynchronous}. The LSTM cell tries to infer the useful information from all previous state representations ($o_{0...t}$) and outputs the hidden state ($h_t$) at the current timestamp as a comprehensive state representation ($s_t$) for the actor- and critic-network (see Fig.~\ref{Fig_A2C}).


\subsection{Multi-Modality State Representation}
\subsubsection{State Embedding from Segmentation}
\par
\revision{The implementation of a UNet} to segment the US image and using it as part of the state representation \revision{is motivated} by the characteristics of RL. The success of RL is based on a large amount of experience, which \revision{necessitates} not only a large amount, but also a diverse set of data~\cite{zhao2020sim}. To implement a trained model into real scenarios, the data ought to be collected from the real scene. In the task of finding the largest longitudinal sections along the vessel, the US data must then be gathered from realistic vascular phantoms, or ideally from patients and volunteers. \revision{Taking into account the distinction} between phantoms and humans tissue and even individual differences between humans, it is hard to transfer the trained model to similar applications without retraining, which is time-consuming~\cite{lake2017building}. However, \revision{by using} a UNet as a preprocessing step, the vascular US images in different applications will have similar geometries, \revision{allowing} the learned model \revision{to be easily transferred to} other similar applications. \revision{It is sufficient to retrain the UNet rather than the entire RL agent because the training time is much shorter. there are already a plethora of mature segmentation techniques}~\cite{weng2019unet,christodoulou2012full}. 
\par
 \revision{Due to} the use of raw images, the size of the features extracted from the images \revision{is} all larger than $256$ in~\cite{hase2020ultrasound,li2021autonomous,ning2021autonomic}. This creates an ample state space, \revision{making} it difficult to train an applicable RL agent. Since the RL algorithms are an experience-driven learning procedure, it is evident that using low-dimensional representations is \revision{preferable to} using high-dimensional ones when both can contain the necessary information~\cite{arulkumaran2017deep}. \revision{By segmenting the US images, it is possible to represent all information from a history image buffer of size $4$ with a feature size of $20$, greatly reducing the complexity of the state space while ensuring good network convergence.}

\subsubsection{Multi-modality State Concatenation}
\par
The feature extractor is modified from SonoNet-16~\cite{baumgartner2017sononet} by deleting the last softmax layer. The UNet structure is identical to that of~\cite{jiang2021autonomous}. Inspired by~\cite{yun2017action}, the history of the actions is also involved in the state representation. Asides from the segmented images and actions history, the state representation also includes information about the area changes of the segmented vessel. When the agent is given information about the area difference between each timestamp, the agent can acquire a better understanding of the environment because the reward is also related to the size of the segmented area. The effect of providing this extra information will be further validated in the experiments.

\begin{figure}[ht!]
\centering
\includegraphics[width=0.48\textwidth]{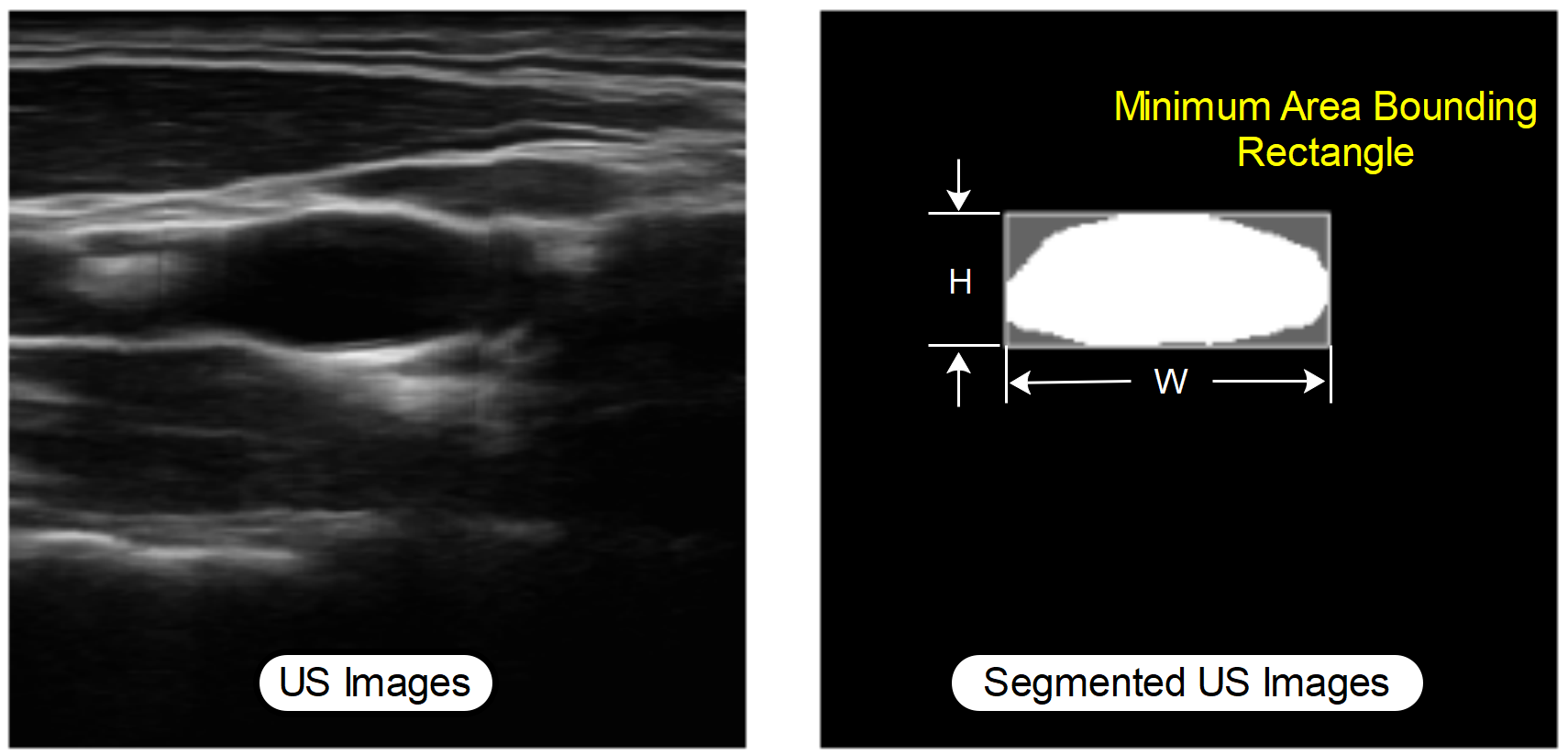}
\caption{Illustration of the minimum area bounding rectangle based standard view recognition method.}
\label{Fig_MABR}
\end{figure}

\subsection{Standard View Recognition}
\par
To terminate the searching process when the largest longitudinal section along the vessel is found, a standard view recognition method is proposed. \revision{Rather than} using a network~\cite{hase2020ultrasound,li2021autonomous}, we applied an approach based on the minimum area bounding rectangle of the vessel in the US image. Compared to a network structure, it is more straightforward and time-efficient. \revision{The segmented vessel's minimum bounding rectangle is calculated} as shown in Fig.~\ref{Fig_MABR}. In each step $t$, the diameter of the vessel $d_v$ is estimated by:

\begin{equation}\label{diameter_estimation}
d_{v,t}=\frac{1}{t}\sum_{k=0}^{t}H_k
\end{equation}
where $H_k$ is the height of the minimum area rectangle in step $k$. Then, a termination ratio $R_{ter}$ is calculated to measure the similarity between the segmented area and the rectangle:
\begin{equation}\label{terminate_ratio}
R_{ter}=\frac{H_t\times W_t-D_t}{H_t\times W_t}
\end{equation}
where $H_t$ and $W_t$ are the height and width of the minimum area rectangle respectively and $D_t$ is the size of the segmented vessel. $H_t\times W_t-D_t$ basically represents the gray area in Fig.~\ref{Fig_MABR}. Since the standard view in our use case is approximately a rectangle, the ratio should be as small as possible.
\par
The whole searching process will be terminated when the following conditions are fulfilled:
\begin{equation}\label{terminate_conditions}
\begin{array}{ll}
Condition1:~&R_{ter}<0.1 \\
Condition2:~&d_{v,t}-H_t<Th \\
Condition3:~&W_t>\alpha W_i 
\end{array}
\end{equation}
where $Th$ is a small threshold set to $10~pixels$ in practice, $W_i$ is the width of the US image in pixel, and $\alpha$ is a discount factor with a value of $0.99$. $Condition1$ is used to ensure that the segmented vessel \revision{resembles} a rectangle \revision{as closely as possible}. $Condition2$ ensures that the height of the bounding rectangle is roughly equal to the vessel's estimated diameter, implying that the US image plane intersects with the centerline. $Condition3$ makes sure that the bounding box's width is equal to the width of the US image. It is given by the characteristic features of the standard view, where the vessel appears as a rectangle across the entire US image.

\begin{table*}
\centering
\caption{Performances of different architectures in virtual environment.}
\label{table_virtual_env}
\begin{threeparttable}
\begin{tabular}{c|c|c|c|c|c|c}
\hline
Method       & Test Environment & Success rate & \makecell{Average number\\of steps} & Position error (mm) & Orientation error ($^\circ$)&  \makecell{Number of\\samples}\\ \hline
\multirow{3}{*}{\textbf{VesNet-RL}}        & Vascular phantom    & $92.3\%$   & $15$   & $2.06\pm1.37$   & $4.08\pm3.20$ & $300$\\ 
                                     & Carotid             & $91.5\%$   & $13$   & $1.29\pm0.83$   & $1.32\pm2.55$ & $400$\\ 
                                     & Carotid$^*$           & $56.5\%$   & $18$   & $0.92\pm0.64$   & $2.25\pm2.75$ & $400$\\ \hline
\multirow{2}{*}{LSTM$^{\spadesuit}$}        & Vascular phantom    & $52.0\%$   & $25$   & $2.06\pm1.22$   & $3.51\pm3.34$ & $300$\\
                                     & Carotid             & $41.3\%$   & $27$   & $1.18\pm0.83$   & $1.77\pm2.85$ & $400$\\ \hline
Segmentation$^{\spadesuit}$                & \makecell{Vascular phantom\\($Vessel_{2\&3}$)}    & $18.7\%$   & $34$   & $1.75\pm1.27$   & $3.33\pm3.26$ & $200$\\ \hline
\multirow{2}{*}{Area changes$^{\spadesuit}$}& Vascular phantom    & $73.0\%$   & $28$   & $1.67\pm1.19$   & $3.22\pm3.11$ & $300$\\
                                     & Carotid             & $64.3\%$   & $27$   & $1.38\pm0.94$   & $2.72\pm3.15$ & $400$\\ \hline
\multirow{2}{*}{Historical information$^{\spadesuit}$} & Vascular phantom    & $52.3\%$   & $32$   & $1.75\pm1.07$   & $3.26\pm3.15$ & $300$\\ 
                                     & Carotid             & $48.5\%$   & $32$   & $1.40\pm0.87$   & $1.71\pm2.82$ & $400$\\ \hline
\multirow{2}{*}{\makecell{VesNet-RL\\(image buffer size: $8$)}} & Vascular phantom    & $24.0\%$   & $25$   & $1.37\pm1.21$   & $3.62\pm3.15$ & $300$\\ 
                                     & Carotid             & $10.3\%$   & $40$   & $1.62\pm0.94$   & $1.72\pm2.83$ & $400$\\ \hline
\end{tabular}
\begin{tablenotes}
        \footnotesize
        \item[$\spadesuit$] means the corresponding module is removed from the proposed VesNet-RL framework.
\end{tablenotes}
\end{threeparttable}
\end{table*}

\section{Experiments and Results}
\subsection{Hardware Setup}
\par
The proposed US standard plane acquisition system is built by two parts: a robotic arm (KUKA LBR iiwa 7 R800, KUKA Roboter GmbH, Augsburg, Germany) controlled using a Robot Operating System (ROS) interface~\cite{hennersperger2016towards} and two different types of US imaging systems. A first (Cephasonics, California, USA) with a linear US probe (CPLA12875, Cephasonics, California, USA) is used to acquire US images from vascular phantoms. A second (ACUSON Juniper Ultrasound System, Siemens AG, Erlangen, Germany), also equiped with a linear US probe (12L3, Siemens AG, Erlangen, Germany), is applied to take US images of human carotids, since it provides qualitatively better human tissue images. The US probes are mounted to the end-effector of the robot by 3D-printed holders. The US settings are mainly adopted from the build-in files from the manufacturers for vascular imaging. The US images from Cephasonics are accessed by a USB interface provided by the manufacturer, while the B-mode images from the Siemens system were captured by a frame grabber (DVI2USB 3.0, Epiphan Video, Ottawa, Canada). The pose of the robot arm is synchronized with the US images in real-time in a software platform (ImFusion Suite, ImFusion GmbH, Munich, Germany) to reconstruct 3D US volumes of the region of interest and build a 3D virtual environment for the US acquisitions.
\par
To validate the performance of \revision{VesNet-RL} in different vascular standard plane searching tasks, three custom-made blood vessel phantoms ($Vessel_1$, $Vessel_2$, and $Vessel_3$) were employed. They were made of gelatin powder ($175~g/L$), paper pulp ($3-5~g/L$), and liquid disinfectant mixed with water, where the paper pulp is used to mimic the human tissue, and liquid disinfectant is adopted to extend its preservation time. To mimic the structure of vessels, after the solidification of the gel, a round tube was used to create holes in different depth of the phantoms.

\subsection{Training Details}

\subsubsection{UNet Training}\label{sec:UNet_training}
\par
The UNet for the vascular phantoms was trained using $4,421$ US images acquired from $Vessel_1$ with various poses of the US probe relative to the vessel. The US images, which only display backgrounds, are included in the training dataset to teach the network \revision{how to recognize the presence} of vessels. If background images are excluded from the training dataset, the performance of the UNet is very unstable when there is no vessel in the images.
\par
The training data for the human carotid UNet consists of $1,041$ US images of a volunteer. The acquisition was performed within the Institutional Review Board Approval by the Ethical Commission of the Technical University of Munich (reference number 244/19 S), having the volunteer signed an informed consent. Considering the carotid pulse during the US sweep, the vascular wall exhibits a wave appearance in the longitudinal view of the vessel in the reconstructed US volumes. As a result, \revision{these data had to be included in the training set}. A sweep along the carotid was performed to reconstruct the US volume of the artery. The probe was attached to the robotic arm, positioned approximately orthogonal to the carotid centerline, \revision{and manually moved along the vessel, with the robot only assisting in the pose acquisition.} 
\revision{The sweep frames were manually labeled after the carotid reconstruction, and another 3D volume of the same size was built using the labeled images.}
\revision{By taking images in the same position in these two compounding volumes, a pair of training data can be collected.}
The training set included $1,266$ images from a 3D compounding volume. In total, $2,307$ images served as the training data for the carotid UNet.

\subsubsection{RL Agent Training}\label{sec:RL_training}
\par
\begin{figure}[htbp]
\centering
\includegraphics[width=0.48\textwidth]{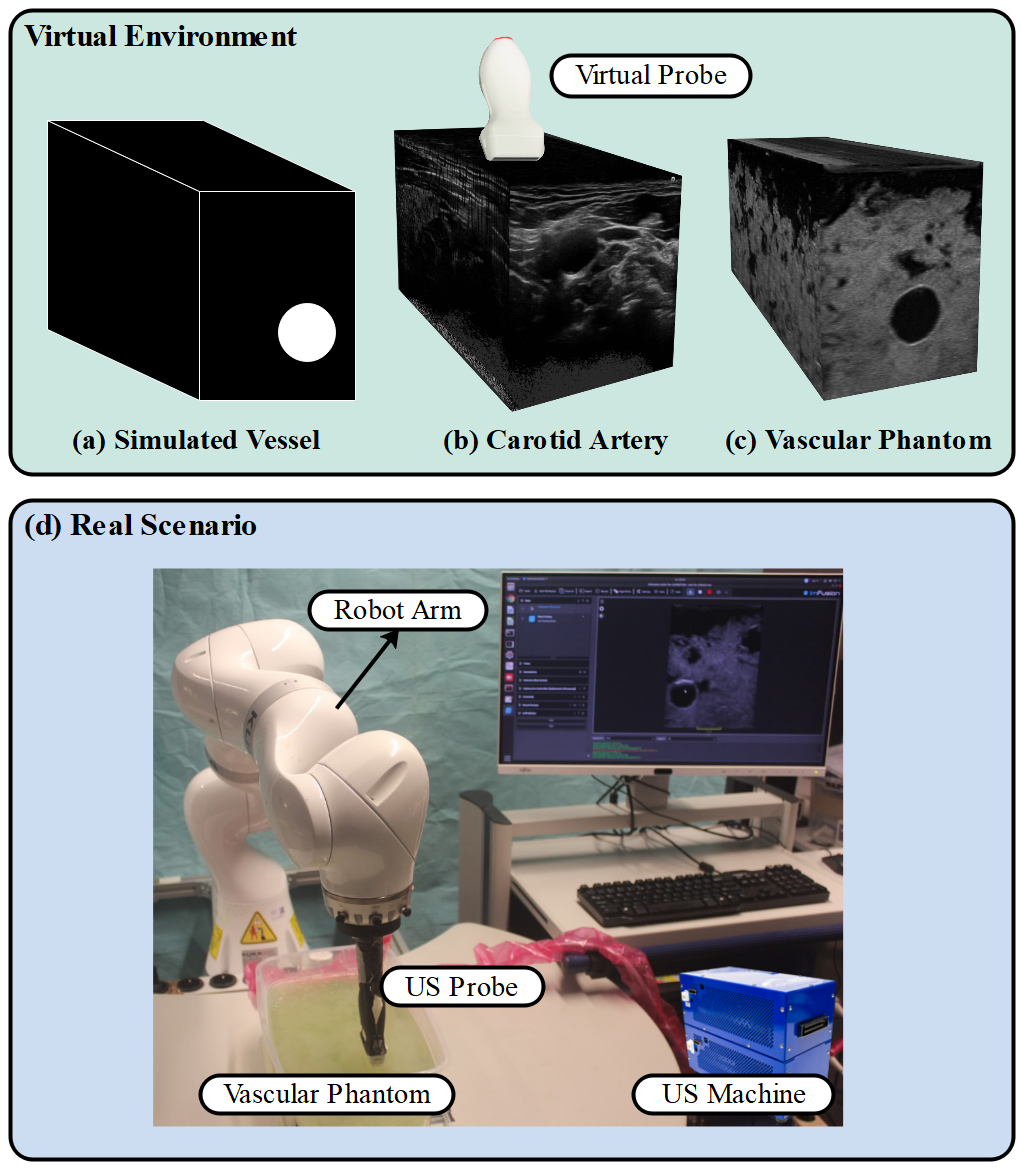}
\caption{Different virtual environments, including (a) simulated binary vessel, (b) carotid artery, and (c) vascular phantom, and (d) the experiment setup in real scenario on a vascular phantom.}
\label{Fig_virtual_env}
\end{figure}

For training of the RL agent, a virtual environment is built on simulated vessels, with the vessels appearing as a white tube on a black background (see Fig.~\ref{Fig_virtual_env}a). The depth and size of the vessels are generated at random. Ten binary vessels are created for the training of the RL agent. There is no need to apply a UNet for the training because the simulated vessel images already have the same characteristics as the segmented US frames. The history image buffer size is set to $4$. The size of the hidden state of the LSTM cell is set to $256$. A total of $3,000$ training episodes were executed. At the beginning of each episode, a vascular environment is randomly selected, and the RL agent is randomly initialized, where the vessel is at least partially observable. The network was trained every 20 interaction steps using Adam optimize, \revision{with a maximum step size of $500$ in each episode}. \revision{The learning rate for the first $500$ episodes is $5\times 10^{-4}$, then drops to $3\times 10^{-4}$ for the next $1,000$ episodes, and finally declines to $1\times 10^{-4}$ for the remaining $1,500$ training episodes.}

\subsection{Experiments in Virtual Environments}\label{sec:virtual_experiment}
\par
In order to demonstrate the generalization ability of VesNet-RL, the RL agent, which was purely trained on simulated vessels, was tested in virtual environments built by the 3D images of vascular phantoms (see Fig.~\ref{Fig_virtual_env}b, Fig.~\ref{Fig_virtual_env}c). In each testing episode, the agent \revision{is randomly initialized in each testing environment} where the vessel is at least partially observable. \revision{The agent is considered successful if it can complete its search in less than 50 steps using the proposed termination criteria.} Furthermore, the efficiency of the trained model is evaluated by calculating the average steps of all successful test episodes. The translational error is defined as the distance between the \revision{probe} tip and the projected centerline of the vessel on OS (equivalent to the upper surface of the US volume). The rotational error is then calculated by the angular difference between the probe and the projected vessel centerline on OS.

\subsubsection{Evaluation of the Standard View Recognition Module}
\par
\revision{The ground truth position of the vessel centerline is used to calculate the position and orientation error in the fifth and sixth columns of Table~\ref{table_virtual_env}.} To obtain the vessel centerline of the vascular phantom in the corresponding 3D images, a sweep along the vessel in the virtual environment is executed, and the centers of the segmented area are fitted into a linear regression model to form a line~\cite{liu2022globally}. As shown in the Table~\ref{table_virtual_env}, our standard view recognition method is able to terminate the searching process with extremely high precision in position and orientation.

\subsubsection{Evaluation of the Architecture Design}
\par
To demonstrate the efficacy of our framework design, we compared \revision{it to various} leave-one-out models (ablation study), in which one of the modules of our original design is removed while the rest remains unchanged. The training details for all the different architectures are the same as in Section~\ref{sec:RL_training}, except the one without segmentation. Because of the absence of the UNet, the simulated vessels cannot be used as training data. Instead, three 3D images of the vascular phantom ($Vessel_1$) are used as the training dataset, and the trained model is then tested on $Vessel_{2\&3}$. The test dataset for the other architectures includes all three vascular phantom models.
\par
\revision{When the LSTM cell is removed from the original design, the success rate drops dramatically, demonstrating} that only considering the previous information in the state representation is \revision{insufficient}. LSTM exhibits superior performance in revealing the underlying persistence in sequential data.
When no segmentation network is employed, the performance of the trained agent shows a weak generalization ability in analogous application environments.
When the area changes information is excluded from the state representation, the success rate drops by~$20\%$, and the average number of steps to the goal \revision{nearly doubles when} compared to VesNet-RL because the area changes information can tell the agent whether it is moving or rotating in the right direction.

\par
\revision{We trained a model that only takes the current observations as state representation to showcase that multiple consecutive images are still required even after the LSTM cell is implemented.}
By comparing the result to the original model, whose state consists of $4$ consecutive images, actions, and area changes, we can conclude that including previous information in the state representation \revision{using an LSTM cell still improves the model}. Because for symmetrical structures like vessels, the same image can be acquired in the same position; but with different orientations, a sequence of consecutive images along with actions history allows the network to gain a better understanding of the surrounding and eliminate the ambiguity, \revision{resulting in more accurate state descriptions and faster training.} \revision{However, when the history images buffer size is set to $8$ and the corresponding images buffer feature size is set to $40$, the trained model performs poorly compared to others, showing that expanding the state space can sometimes prevent models from learning a delicate policy.}






\subsubsection{Performance Comparison between Phantom and In-Vivo Human Data}
\par
To test our model in a more realistic scenario, four 3D models of human carotid were built as described in Section~\ref{sec:UNet_training} (see Fig.~\ref{Fig_virtual_env}). It is worth noting that the US machine used for human data acquisition differs from the one used for vascular phantoms. , Carotid$^*$ in Table~\ref{table_virtual_env} indicates that the images from the 3D reconstructed volume are not included in the UNet training set for carotid as described in Section~\ref{sec:UNet_training}.
If the UNet fails to segment the vessel area properly, then the success rate of the RL model is much lower. On the contrary, when images from the 3D reconstructed volume of the carotid are included in the UNet training data, \revision{VesNet-RL achieves a $91.5\%$ success rate in locating the longitudinal section of the carotid.} For the other architectures, the success rates drop slightly. Except for the one without segmentation, if there is no retraining, the trained network is not able to be transferred to carotid applications.

\begin{table}[htbp]
\centering
 \caption{Performance of \revision{VesNet-RL} in real scenario}
 \label{Table_real_scenario}
 \begin{tabular}{cccc}
  \toprule
   Success rate & \makecell{Average number\\of steps} & \makecell{Position\\error (mm)} & \makecell{Orientation\\error ($^\circ$)}\\
  \midrule
$80.0\%$   & $17$   & $0.79\pm0.55$   & $2.08\pm3.05$\\
  \bottomrule
 \end{tabular}
\end{table}

\subsection{Experiments in Real Scenarios}
To showcase the performance of VesNet-RL in real scenarios, we tested our trained model on the real vascular phantoms with a robot arm (see Fig.~\ref{Fig_virtual_env}). \revision{The trained model is the same as in Section~\ref{sec:RL_training}. The OS is then defined as the upper surface of the gel phantom, while the actions are identical to Section~\ref{sec:Action_Space}. The phantom is immerged into water so there is no need to use US gel.} $60$ tests were carried out on a custom-made vascular phantom. The robot executed the learned policy with a maximum of $50$ steps. At the beginning of each test, the probe is randomly initialized orthogonal to the upper surface of the phantom, where the vessel is at least partially observable. The defination of success rate is identical to that of Section~\ref{sec:virtual_experiment}. Table~\ref{Table_real_scenario} shows that \revision{VesNet-RL} \revision{has a high success rate ($80\%$) and high accuracy in navigating the US probe to the standard view of a vessel.}
\revision{In the vast majority of cases, the failure was due to incorrect vessel segmentation, which resulted in a misestimation of the actual state.}


\subsection{Discussion}
\par
\revision{
Besides the anatomy of interest, the background of US images also contains certain information. For the tasks like searching for specific anatomies, e.g., kidney, the background can also help clinicians quickly locate the anatomy of interest, particularly when the searching process starts from a random position. However, for tasks like locating the standard planes of arteries (e.g., longitudinal view), the displayed view of the objects is more important to accurately navigate the probe. Since the background of B-mode images is sensitive to practical factors like contact force, amount of gel, and orientation, which will hinder the convergence of the trained model and affect the generalizability of the trained model for unseen patients.
}


\section{Conclusion}
In this work, we present a simulation-based RL approach for automatically navigating a US probe to a vascular standard plane (i.e., the largest longitudinal view).
\revision{Segmented binary images are used as part of a multimodality state representation to bridge the gap between the simulation training environment and the real scenario, as well as to address the challenge of low generalization ability}
Thanks to an explicit segmentation of the US frames, the RL agent, trained with a wide \revision{variety} of simulated binary vessels, can be used to guide the US probe in actual practice, such as the carotid standard view acquisition.
\revision{Experiments were conducted in both virtual and real scenarios to demonstrate the efficacy of VesNet-RL. The proposed model was compared with various network structures on 3D models of vascular phantoms and a human carotid in virtual environments. With the highest success rate ($92.3\%$ for vascular phantoms and $91.5\%$ for the carotid artery) and the minimum average number of steps ($15$ for vascular phantoms and $13$ for the carotid artery), the proposed framework outperforms the competition.} The novel standard view recognition method for vascular use-case also achieves excellent results in the tests for tubular phantom and carotid artery in the virtual environments ($2.06\pm1.37~mm$ and $1.29\pm0.83~mm$ in terms of position error and $4.08\pm3.20^\circ$ and $1.32\pm2.55^\circ$ in terms of orientation error, respectively). We also demonstrate that the model trained in a simulation environment can be directly applied in the real scenario on a vascular phantom without extra training or retraining, achieving a success rate of $80\%$, $0.79\pm0.55~mm$ position error, and $2.08\pm3.05^\circ$ orientation error. In the future, we will further consider the contact force and the deformation of the human tissue~\cite{jiang2021deformation} to further pave the way to real clinical applications.

\bibliographystyle{IEEEtran}
\bibliography{IEEEabrv,references}

\end{document}